\documentclass[12pt,preprint]{revtex4-1}
\usepackage{graphicx}
\usepackage{amsmath}
\usepackage{amssymb}
\usepackage{color}
\usepackage{ulem}
\usepackage{}

\begin{document}
\title{Tomography of asymmetric molecular orbitals with one-color inhomogeneous field}

\author{Hua Yuan, Lixin He,$^{1,}$\footnote{helx@mail.hust.edu.cn} Feng Wang,$^{1}$ Baoning Wang,$^{1}$ Xiaosong Zhu,$^{1}$ Pengfei Lan,$^{1,}$\footnote{pengfeilan@mail.hust.edu.cn} and Peixiang Lu$^{1,2,}$\footnote{lupeixiang@mail.hust.edu.cn}
}

\affiliation{%
 $^1$School of Physics and Wuhan National Laboratory for Optoelectronics, Huazhong University of Science and Technology, Wuhan 430074, China\\
$^2$Laboratory of Optical Information Technology, Wuhan Institute of Technology, Wuhan 430205, China
}%



\begin{abstract}
We demonstrate to image asymmetric molecular orbitals via high-order harmonic generation in a one-color inhomogeneous field. Due to the broken inversion symmetry of the inhomogeneous field in space, the returning electrons with energy in a broad range can be forced to recollide from only one direction for all the orientation angles of molecules, which therefore can be used to reconstruct asymmetric molecular orbitals. Following the procedure of molecular orbital tomography, the highest occupied molecular orbital of CO is satisfactorily reconstructed with high-order harmonic spectra driven by the inhomogeneous field. This scheme is helpful to relax the requirement of laser conditions and also applicable to other asymmetric molecules.
\end{abstract}

\maketitle

In past years, high-order harmonic generation (HHG) through the nonlinear interaction between an intense laser field and atoms or molecules has attracted considerable theoretical and experimental attention. This interest in HHG arises from its important applications. For example, using HHG, one can obtain attosecond pulses \cite{J. Li,T. Gaumnitz}, which offer robust tools for probing and controlling ultrafast electronic dynamics inside atoms \cite{E. Goulielmakis}, molecules \cite{S. Baker,P. Lan}, and solids \cite{G. Vampa,M. Schultze,X. Liu1}. In addition, the emitted high-order harmonics in molecules contains a wealth of information about the structure of its generating medium. This has stimulated molecular orbital tomography (MOT) \cite{J. Itatani,YangLi,C. Zhai,C. Vozzi,C. Zhai1,E. V. van der Zwan,Y. J. Chen,M. Qin,S. A. Rezvani,B. Wang,Z. Hong}, which provides potential applications in observing femtosecond electron dynamics in chemical reactions \cite{H. J. W,P. Sali}. The MOT was first proposed and demonstrated by Itatani \textit{et al.} for imaging the highest occupied molecular orbital (HOMO) of $N_2$ \cite{J. Itatani}. Afterwards, it has been extended to more complex molecules, like $CO_2$ \cite{C. Vozzi} and $C_2H_2$ \cite{C. Zhai1}. Recently, much attention has been paid to the asymmetric molecules such as CO \cite{E. V. van der Zwan,M. Qin,Y. J. Chen,B. Wang}. Nevertheless, the original procedure of MOT encounters difficulty for asymmetric molecules, since the MOT of asymmetric molecules usually requires the control of unidirectional recollision of the electron wave packets. It has been reported that the unidirectional recollision of electrons can be achieved by using extremely short phase-stabilized laser pulses (single-cycle) \cite{E. V. van der Zwan}. However, the requirement of these pulses is rather stringent for many laboratories. Then other methods, like using a two-color laser field and a two-color orthogonally polarized laser field have also been proposed \cite{M. Qin,B. Wang}. But these schemes rely on the control and stabilization of the carrier-envelope phase (CEP) and relative phase of the laser pulses. To further relax the requirement, more simple methods with a one-color driving laser pulse for MOT of asymmetric molecules are still desired.

Recently, HHG in the vicinity of nanostructures has attracted much attention. Due to the surface plasmon resonances within metallic nanostructures, the intensity of the incident laser field can be enhanced by several orders of magnitude \cite{S. Kim,M.sivis,S.Kim1,I. Y. Park,M.Sivis1}. The enhanced laser intensity easily exceeds the threshold intensity for HHG in noble gases. By using the bow-tie-shaped nanostructures, Kim et al. \cite{S. Kim} first observed the plasmon-driven HHG in experiment. However, the outcome of Kim's experiment has been subject to an intense controversy due to the inefficient harmonic emissions which will be overpowered by plasma atomic lines \cite{M.sivis,S.Kim1}. Afterwards, the plasmon-driven HHG has been demonstrated in some other nanostructures \cite{I. Y. Park,M.Sivis1}. In the nanogap where HHG takes place, the enhanced field is spatially inhomogeneous. Driven by such an inhomogeneous field, the electron dynamics can be effectively controlled and the HHG in the nanostructures shows some novel characteristics \cite{MFCiappina,B. Fetic,B. Fetic1,J.A.P,L. He,I. Yavuz,F. Wang,M. F. Ciappina,J. Luo,T. Shaaran,H. Yuan1,ZheWang}, for example, the generation of even order harmonics, the selection of the quantum path and the extension of the harmonic cutoff. It has been demonstrated that the limited spectral range in homogeneous fields will give raise to some artificial structures in the reconstructed orbital \cite{J. Itatani}. In inhomogeneous fields, the harmonic cutoff extension permits denser sampling in the spatial frequency domain, which therefore would be more advantageous for MOT.

In this paper, we demonstrate to reconstruct asymmetric molecular orbitals based on HHG in a one-color inhomogeneous laser field. By using the inhomogeneous laser field, the returning electrons with the energy in a broad range can be efficiently controlled to recollide from only one direction when molecules are oriented at different angles. This permits the MOT for asymmetric molecules. With the harmonic spectra generated in the inhomogeneous field, the HOMO of CO is well reconstructed. The method using the one-color inhomogeneous field to achieve MOT, does not need an ultrashort duration or a stable CEP of the driving pulse, which can extremely relax the requirement of laser conditions.

\begin{figure}[htbp]
	\centering
	\fbox{\includegraphics[width=10cm]{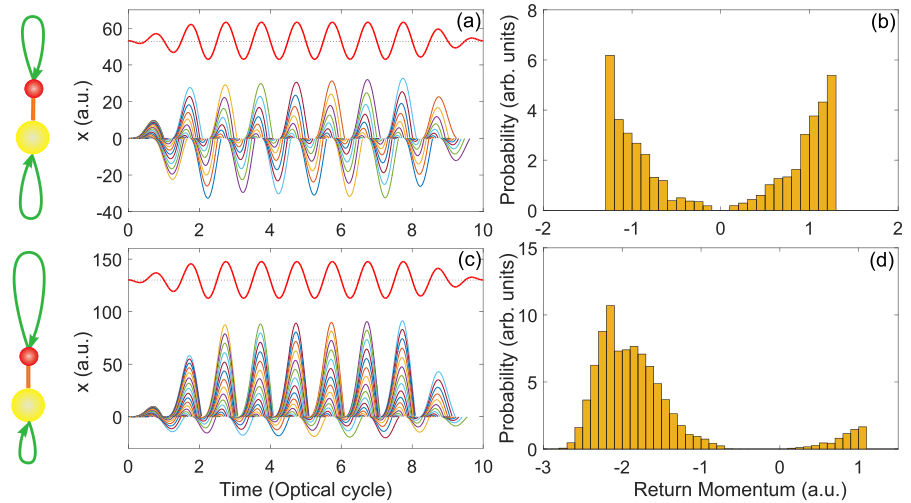}}
	\caption{(Color online) (a)-(b) The classical electron trajectories and returning electron momentum distribution in the one-color homogeneous field. (c)-(d) same as (a)-(b), but for the case of inhomogeneous field. The inserted red solid lines in (a) and (c) are the electric fields. On the left side of (a) and (c), the sketches of the returning electrons in the one-color homogeneous and inhomogeneous fields are shown. Here, the CO molecule is oriented at $0^\circ$. }
	\label{fig:false-color}
\end{figure}

We first demonstrate to control the electron dynamics by using a one-color inhomogeneous laser field. In our simulations, the spatial-dependent laser field is linearly polarized along the $\vec{x}$ direction, which is given by ${E(x,t) = E_t(t)(1 + \varepsilon x)}$ \cite{I. Yavuz,J. Luo,H. Yuan1,ZheWang}. $x$ is the position of the electron ($x=0$ refers to the excursion of the parent ion). The parameter $\varepsilon$ determines the strength of the spatial inhomogeneity of the laser field, $\varepsilon=0$ corresponds to the homogeneous field. The electric field $E_t(t)$ is defined by ${E_t(t)} = {{E_0}f(t)\sin [{\omega _0}t + {\phi _0}]}$ with $E_0$, $\omega _0$ and $\phi _0$ the amplitude, angular frequency and CEP of the laser field, respectively. Here, we use an 800-nm laser pulse with the laser intensity $I_0$ of $1.2\times10^{14}\ \mathrm{W/cm}^2$. The CEP $\phi _0$ is chosen to be 0. $f(t)$ is a trapezoidal envelope and the total pulse lasts for ten optical cycles with two-cycle turn-on and turn-off. The electric field is inserted in Figs. 1(a) and 1(c) with the red solid line. Based on the three-step model \cite{P. B. Corkum}, we have calculated the classical electron trajectories in the inhomogeneous field, as shown in Fig. 1(c). Here, the inhomogeneity parameter $\varepsilon$ is 0.01. For comparison, the result in the one-color homogeneous field is also plotted in Fig. 1(a). One can see that in the homogeneous field [Fig. 1(a)], the electrons ionized in the adjacent two half cycles move towards two opposite directions (the negative-$x$ and positive-$x$ directions). Due to the inversion symmetry, the largest excursion distances of electrons in both directions are nearly the same. Then the maximum momentum values of the returning electrons from these two directions should be almost equal. To confirm this, we have calculated semi-classically the probability that an electron returns to the core with momentum k. The return probability is depicted based on two factors \cite{E. V. van der Zwan1}. One is the tunnel ionization rate at the ionization time $t_i$, which is calculated by the Molecular Ammosov-Delone-Krainov model for oriented molecules \cite{X. M. Tong}. The other factor is $\tau^{-3}$, where $\tau$ is the time the electron spends in the continuum until the time of return. This factor reflects the effects of wave-packet spreading \cite{M. Lewenstein}. The calculated result in the homogeneous field is presented in Fig. 1(b). From this figure, one can see that the maximum positive and negative momenta of the returning electrons are k $\simeq$ $\pm$1.3 a.u.. As a consequence, in the homogeneous field, the generation of each harmonic is attributed to the recollision of returning electrons from both the negative-$x$ and positive-$x$ directions. While in the inhomogeneous field [see Fig. 1(c)], the electrons ionized around the negative peaks of the electric field leave towards the positive-$x$ direction, and finally return to the parent ion with negative momenta. Since the electrons are accelerated by the electric field $E(x,t) = E_t(t)(1 + \varepsilon x)$ whose effective peak amplitude increases along the positive-$x$ direction, the electrons move farther away from the parent ion and gain more kinetic energies \cite{I. Yavuz,H. Yuan1,ZheWang}. In this case, the maximum negative momentum of the returning electrons is about -2.9 a.u. [Fig. 1(d)], which is larger than that in the homogeneous case. This means a cutoff extension in the inhomogeneous field. On the contrary, the electrons ionized around the positive peaks of the electric field leave towards the negative-$x$ direction, and finally return to the parent ion with positive momenta. In the negative-$x$ direction, the effective peak amplitude of the laser field decreases with the electron excursion. The largest excursion distances of electrons are obviously suppressed. The maximum momentum of the returning electrons from the negative-$x$ direction is only about 1.1 a.u. [see Fig. 1(d)], which is much smaller than that from the positive-$x$ direction. Due to the breaking of the inversion symmetry in the inhomogeneous field, the returning electrons with the momentum value above 1.1 a.u. are effectively controlled to recollide from only one direction (the positive-$x$ direction).

To confirm the above analyses, we have calculated the harmonic spectra and the corresponding time-frequency distributions driven by these pulses. Figures 2(a) and 2(b) show the results of CO molecule in the one-color homogeneous field with the orientation angle $\theta=0^\circ$ (the angle between the molecular axis and the laser polarization direction). Here, $\theta=0^\circ$ means that the molecule is oriented such that the largest electric field oscillation points from C to O. The harmonic spectrum is calculated by solving the two-dimensional time-dependent Schr\"{o}dinger equation (2D-TDSE) under the sigle-active-electron approximation with the second-order split-operator method \cite{M. D. Feit}. We use the soft-core potential $V(r) =  - \frac{{({Z_{1i}} - {Z_{1o}}){e^{- ({(r - {R_1})^2}/\rho )}} + {Z_{1o}}}}{{\sqrt {\xi  + (r-R_1)^2} }} - \frac{{({Z_{2i}} - {Z_{2o}}){e^{- ({(r - {R_2})^2}/\rho )}} + {Z_{2o}}}}{{\sqrt {\xi  + (r-R_2)^2} }}$ \cite{Y. J. Chen}. Here $Z_{1i}$=6, $Z_{1o}$=0.6 and $Z_{2i}$=4, $Z_{2o}$=0.4 are the screened effective nuclear charges for the O center and the C center. The subscripts $i$ and $o$ denote the inner and outer limits of $Z_1$ and $Z_2$. $R_1$ and $R_2$ are the positions of the nuclei. $\xi$=0.5 and $\rho$=1/1.746 are the softening and the screening parameters. The ground state is obtained by imaginary time propagation with this soft-core potential and the calculated ground ionization potential is 14 eV, agreeing well with the true value of CO. From Fig. 2(b), one can see that within each optical cycle, there are two emission peaks (marked as $P_1$ and $P_2$) contributing to HHG, which just correspond to the recollision of electrons from both the negative-$x$ and positive-$x$ directions.
\begin{figure}[htbp]
	\centering
	\fbox{\includegraphics[width=10cm]{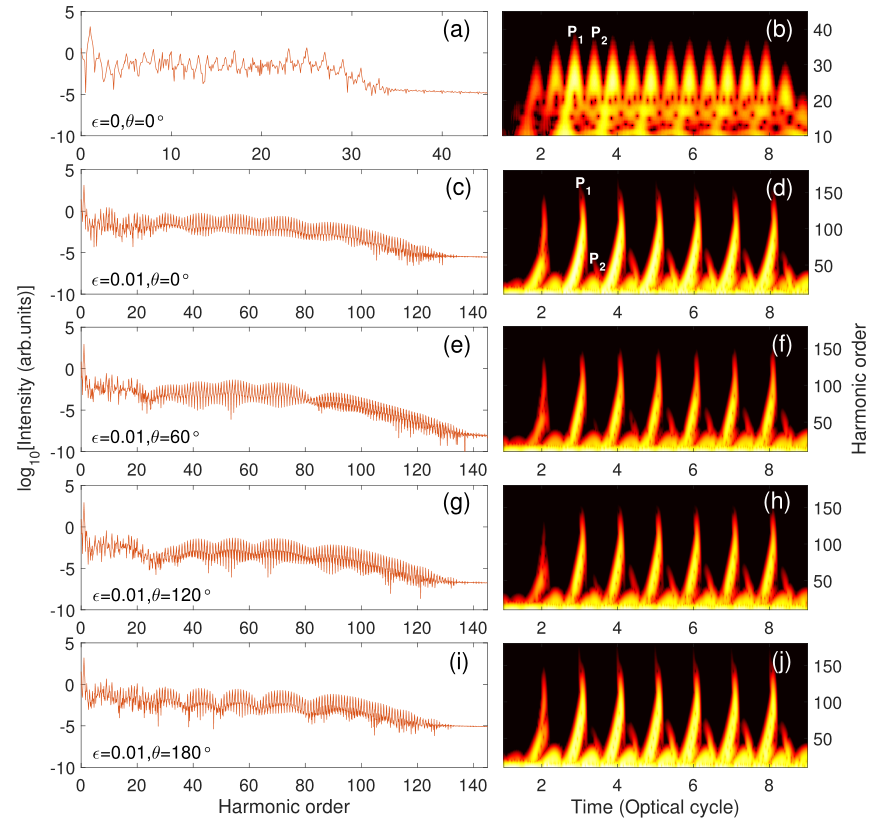}}
	\caption{(Color online) (a) The generated harmonic spectra in the homogeneous field with the CO molecule oriented at $0^o$. (b) The corresponding time-frequency distribution. (c)-(d), (e)-(f), (g)-(h), (i)-(j) same as (a)-(b), but for the cases of inhomogeneous field with the CO molecule oriented at $0^\circ$, $60^\circ$, $120^\circ$ and $180^\circ$, respectively.}
	\label{fig2:false-color}
\end{figure}

The maximum values of two emission peaks are at 25th harmonic, which corresponds to the maximum (positive and negative) momenta of the returning electrons, i.e., $\pm$1.3 a.u.. While in the inhomogeneous field [see Fig. 2(d)], the emission peak $P_1$ is efficiently extended to the 83rd harmonic, corresponding to the maximum momentum of the returning electrons from the positive-$x$ direction, i.e., -2.9 a.u.. The other emission peak $P_2$ is obviously suppressed to the 21st harmonic, which corresponds to the maximum momentum of the returning electrons from the negative-$x$ direction, i.e., 1.1a.u.. As a consequence, in the inhomogeneous field, the high-order harmonics from 21st to 83rd are attributed to the returning electrons from only the positive-$x$ direction. For the MOT of asymmetric molecules, the unidirectional recollision is required for all the orientation angles of molecules. Then we have calculated the harmonic spectra and the corresponding time-frequency distributions in the inhomogeneous field with the CO molecule oriented at different angles. As presented in Figs. 2(d) to 2(j), for all the orientation angles, the unidirectional recollision of the returning electrons is still satisfied for the generation of harmonics in the range from 21st to 83rd, which offers the possibility for the reconstruction of the asymmetric molecular orbital of CO.

\begin{figure}[htbp]
	\centering
	\fbox{\includegraphics[width=10cm]{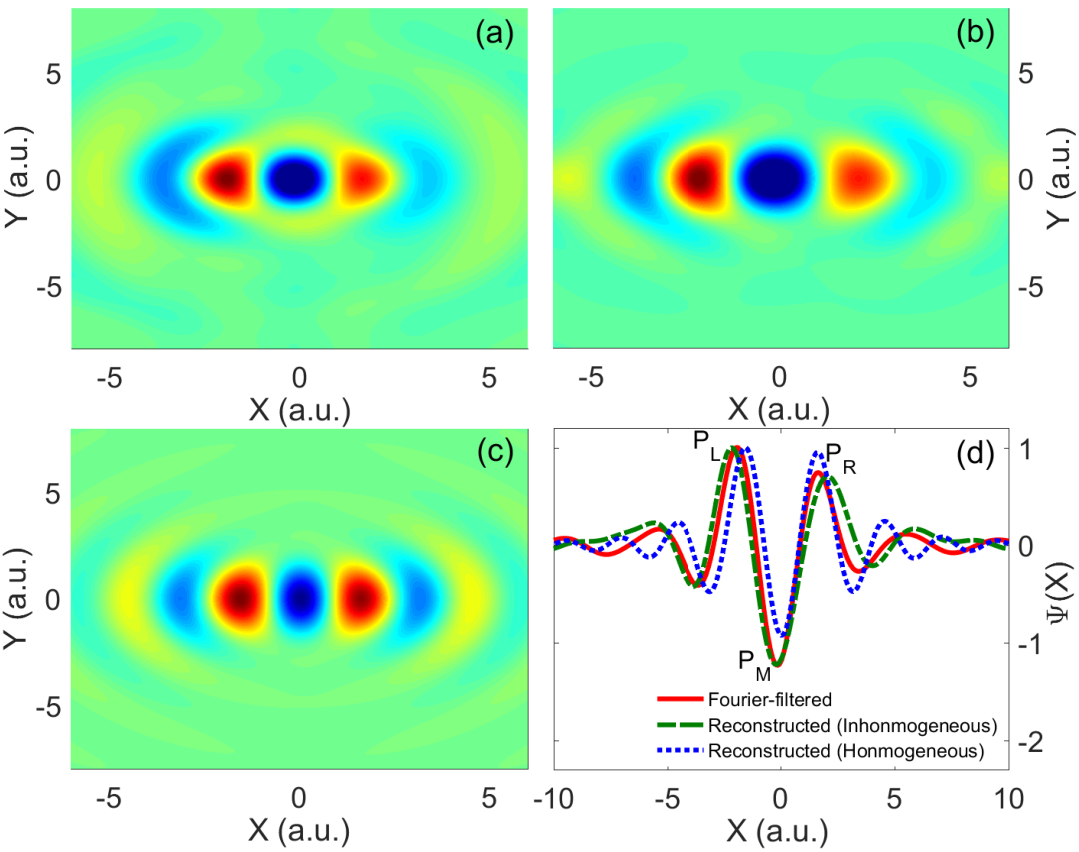}}
	\caption{(Color online) (a) The Fourier-filtered ab initio orbital of the CO molecule. (b) The reconstructed orbital of CO by using the one-color inhomogeneous field. (c) The reconstructed orbital of CO by using the one-color homogeneous field. (d) The cuts along the internuclear axis for the orbitals in (a)-(c).}
	\label{fig3:false-color}
\end{figure}

In the following, we demonstrate to reconstruct the HOMO of CO molecule with harmonic spectra generated in the one-color inhomogeneous field. According to the Fourier slice theorem, the harmonic spectra with molecule oriented at different angles are needed to accomplish MOT \cite{J. Itatani}. To this end, we have calculated the harmonic spectra by solving the 2D-TDSE with orientation angles of CO molecule varying form $0^\circ$ to $360^\circ$ with a step of $\bigtriangleup$$\theta$=${5^\circ}$.  According to the strong-fild approximation (SFA) theory \cite{M. Lewenstein}, the induced dipole moment for HHG at frequency $w$ can be expressed as a factorized expression ${D(w,\theta)=a_{ion}(w,\theta)a_{prop}(w)d(w,\theta)}$. The first two factors $a_{ion}$ and $a_{prop}$ represent the amplitudes of tunneling ionization and  propagation of the recolliding electron wave packet in the continuum, respectively. $d(w,\theta)$ is the recombination dipole moment between the initial orbital and the continuum wave function. The first two factors can be determined by calculating the harmonic spectrum of a reference atom that has the same ionization potential as the molecule (it is Kr for CO). Then in the plane-wave approximation, the HOMO of the molecule can be directly reconstructed by performing inverse Fourier transform of the recombination dipole moment $d(w,\theta)$ \cite{J. Itatani,YangLi,E. V. van der Zwan,E. V. van der Zwan1}. In Fig. 3, we use the odd harmonics ranging from 31st to 81st in the harmonic plateau to do the reconstruction. In Fig. 3(a), we first present the Fourier-filtered  ab initio (FFABI) orbital as a benchmark. To obtain the FFABI orbital, we first calculate the recombination dipole moment $d_{ab}(k)$ by using the ab initio orbital $\psi_{ab}(k,r)$ in terms of $\left\langle \psi_{ab}(k,r)\left| r\right| e^{ikr}\right\rangle $. Then, the dipole $d_{ab}(k)$ is filtered for the same harmonic spectral range and orientation angles as in our reconstruction procedure. Finally, the FFABI orbital is obtained by performing the inverse Fourier transform of the filtered dipole. Figure 3(b) show the reconstructed molecular orbital of CO using the inhomogeneous field. Comparing Fig. 3(b) with Fig. 3(a), one can see that the reconstructed orbital shows good agreement with the FFABI result. In detail, there are three main lobes with the alternating positive and negative signs, separating by two nodal surfaces along the y direction. And the left and right main lobes show obvious difference, corresponding to the asymmetric structure of CO. Whereas, some additional ambient structures, which do not exist in the exact HOMO of the CO molecule, appear in Figs. 3(a) and 3(b). This is due to the limited bandwidth of the harmonic spectra used in our reconstructions \cite{C. Vozzi,M. Qin,B. Wang}. For comparison, we also reconstruct the molecular orbital with HHG in the homogeneous field. In order to guarantee the same bandwidth of the harmonic spectrum, the laser intensity of the homogeneous field is raised to $7\times10^{14}\ \mathrm{W/cm}^2$. The reconstructed orbital is presented in Fig. 3(c). One can see that the left and right main lobes are almost the same, which fails to reproduce the asymmetric structure of the CO molecule. To further verify the quality of the reconstruction in the inhomogeneous field, the cuts along the internuclear axis ($y=0$) for above three orbitals are depicted in Fig. 3(d). It is shown that the two positive maxima ($P_L$ and $P_R$) and the negative maximum ($P_M$) of the reconstructed orbital in the inhomogeneous field (the green dashed line) agree well with that of the FFABI orbital (the red solid line). Moreover, $P_L$ is larger than $P_R$. While, in the homogeneous field (the blue dotted line), the negative maximum $P_M$ and the right positive maximum $P_R$ show obvious difference from those of the FFABI orbital. Even worse, $P_L$ and $P_R$ are nearly comparable, which is inconformity to the exact HOMO of the CO molecule.

In summary, we have theoretically demonstrated to image the asymmetric molecular orbitals based on the HHG in a one-color inhomogeneous field. Under the control of the spatial-dependent laser field, the momentum of the returning electrons from the positive-$x$ direction is extended to -2.9 a.u., while that from the negative-$x$ direction is suppressed to 1.1 a.u.. As a consequence, the unidirectional recollision of the electron wave packet is achieved for HHG in a broad range from 31st to 81st, which therefore can be used for MOT of asymmetric molecules. With HHG in the inhomogeneous laser field, the highest occupied molecular orbital of CO has been successfully reconstructed, which shows quantitative agreement with the Fourier-filtered ab initio orbital. In this scheme, the unidirectional recollision of the electron wave packet benefits from the asymmetry of the inhomogeneous field in space. It doesn't require an ultrashort duration or a stable CEP of the driving laser field. Therefore, our scheme is helpful to relax the requirement of laser conditions for MOT of asymmetric molecules. Moreover, our approach can also be extended to some other asymmetric molecules. It is also worth mentioning that the harmonic cutoff extension in spatial inhomogeneous field will enrich the sampling in the spatial frequency domain, which is helpful to improve the accuracy of the reconstruction and therefore is more advantageous for MOT of both asymmetric and symmetric molecules.

\textbf{Funding.} National Natural Science Foundation of China (NNSF) (11627809, 61275126, 11404123, 11422435, 11234004, 11704137); China Postdoctoral Science Foundation funded Project (2017M610467); Program for HUST Academic Frontier Youth Team.

\end{document}